\documentclass[psfig]{aa}

\usepackage{times}
\usepackage{graphics}
\usepackage{latexsym}
\usepackage{epsfig}

\begin{document}
\headnote{Letter to the Editor}

\title{
Mid-infrared selection of AGN
\thanks{Based on
  observations with the Infrared Space Observatory
  ISO, an ESA project with instruments
  funded by ESA Member States (especially the PI countries: France,
  Germany, the Netherlands and UK) and
  with the participation of ISAS and NASA, and also based on
  observations at the
  South African Astronomical Observatory SAAO.}
}
\author{M.\ Haas\inst{1}
   \and R.\ Siebenmorgen\inst{2}
   \and C.\ Leipski\inst{1}
   \and S.\ Ott\inst{3}
   \and B. Cunow\inst{4}
   \and H. Meusinger\inst{5}
   \and \\
   S.A.H.\ M\"uller\inst{1}
   \and R.\ Chini\inst{1}
   \and N.\ Schartel\inst{6}
\\
}
\offprints{Martin Haas (haas@astro.rub.de)}
\institute{
Astronomisches Institut, Ruhr-Universit\"at Bochum (AIRUB),
Universit\"atsstr. 150 / NA7, 44780 Bochum, Germany
\and
European Southern Observatory (ESO), Karl-Schwarzschild-Str. 2, 85748 Garching, Germany
\and
HERSCHEL Science Centre,
ESA, Noordwijk, P. O. Box 299, 2200 AG Noordwijk,
The Netherlands
\and
Department of Mathematics, Appl. Math. \& Astronomy,
University of South Africa,
PO Box 392, Pretoria 0003, South Africa
\and
Th\"uringer Landessternwarte Tautenburg, Sternwarte 5, 07778 Tautenburg, Germany
\and
XMM-Newton Science Operation Center, ESA, Vilspa, Apartado 50727, 28080 Madrid, Spain
}
\date{Received 26. February 2004; accepted 15. April 2004 }
\authorrunning{M. Haas et al.}
\titlerunning{Mid-infrared selection of AGN}

\abstract{ Since a large fraction of active galactic nuclei (AGN) is
missed in common UV-excess surveys and is even hard to find in radio,
near-IR and X-ray surveys, we have used a new AGN selection
technique which is expected to be
not affected by extinction. Within the scientific
verification of the {\it ISOCAM Parallel Survey} at $6.7\,\mu$m we
have discovered objects with exceptional mid-infrared (MIR)
emission. They are essentially not detected on IRAS-ADDSCANs and only
very few of them show up in the NVSS and FIRST radio
surveys. Various colour criteria of the $6.7\,\mu$m data with 2MASS
and optical wavebands show that the sources 
reach more extreme IR colours than the sources in the Hubble
Deep Field-South and the ELAIS survey. The comparison with
known object types suggests that we have found AGN with a pronounced
MIR emission, probably due to circum-nuclear dust. First
results from optical spectroscopy of ten candidates corroborate this
interpretation showing four AGN, two reddened LINER and four extremely
reddened emission-line galaxies with MIR/FIR flux ratios higher than
for known pure starburst galaxies. The results will make a significant
contribution to the debate on the entire AGN population.

} \maketitle

\section{Introduction}
\label{section_introduction}

The realisation that obscuration plays a critical role in the
classification of AGN fundamentally inspired the current research.
Attempts to overcome the limitations of dust extinction and to
identify the entire AGN population -- including type 2 and
dust-enshrouded AGN -- encompass surveys in the near-IR, radio, and
X-ray regimes. However, searching for very red AGN
the colour selection via $J - K_{\rm s} > 2$  (Cutri
et al. 2001) excludes most of the known AGN (Barkhouse and Hall 2001),
only about 30\% of AGN are radio-loud (Urry \& Padovani 1995), and
there seems to exist many X-ray faint AGN (Wilkes et al. 2002). Thus, a
considerable fraction of the AGN population must have escaped
detection due to observational bias.
Webster et al. (1995) found that their radio-selected quasar sample is
significantly redder than an optical comparison sample and concluded that up
to 80\% of the quasars have been missed in conventional optical
surveys, provided that
the redder colours of the radio-loud quasars are due to dust reddening and
that the radio-quiet quasars contain as much dust as the radio-loud ones.
By new strategies in the optical,
assuming that the narrow-line regions are sufficiently extended and
that only the continuum emission is hidden, type-2 quasar candidates have been
selected as objects with narrow permitted emission lines and
high [\ion{O}{III}]$\lambda$5007 equivalent widths (Djorgovski et al. 2001,
Zakamska et al. 2003).
Applying a moderate colour cut $J-K_{\rm s} > 1.2$, from about 1500 sources
in 2MASS Francis et al. (2004) find only tentative evidence that
Seyfert\,2 nuclei are more common in the NIR selected survey than in
blue selected galaxy surveys, and they can place only very weak
constraints on any population of dusty AGN.

The disadvantage of heavy extinction in optical and NIR surveys
can turn into a valuable detection tool, when observing
dust-surrounded AGN at MIR wavelengths. There, the reemission of
the hiding dust heated by the strong radiation field of the AGN
should be seen easily as MIR excess.
In fact, for known (powerful) AGN of both
type 1 and type 2 a steep near- to mid-IR slope has been revealed
by sensitive MIR observations (e.g. Haas et al. 2003,  Haas et al.
2004, Siebenmorgen et al. 2004).

We therefore started a new approach, searching for AGN by means of
their MIR emission of the nuclear dust torus.
However, one complication with this method has to be solved: Since
luminous IR starburst galaxies may also show a pronounced MIR emission
due to the PAH emission bands around $7.7\,\mu$m, it is of special
importance to distinguish them from AGN. In this Letter we describe the new
technique for the mid-IR selection of AGN candidates using IR
colour diagrams and report about
first results from optical spectroscopy.

\vspace{-0.2cm}

\section{Selection of MIR sources}
\label{section_iso_list}

ISO has performed a serendipitous survey at $6.7\,\mu$m ($LW2$ band),
the {\it ISOCAM Parallel Survey}, with 6$\arcsec$ spatial resolution and a
positional accuracy of better than 3$\arcsec$ (Siebenmorgen et al. 1996, Ott
et al. 2003). Over 27 square degrees of the sky are processed and
currently being catalogued. For point sources the $3\,\sigma$
detection limit is about 0.5\,mJy.

Within the scientific verification of the 17000 detected sources
(Ott et al. 2004 in prep.) we have
selected unresolved sources at galactic latitude
$|b| > 20^\circ$. We then
performed cross correlations with the 2MASS all sky point source
catalogue (Cutri et al. 2003), with the USNO-B, DSS and UCAC optical
catalogues, as well as the NVSS and FIRST radio surveys, and analysed
IRAS ADDSCANs.  We excluded objects which have multiple NIR and
optical counterparts within 10$\arcsec$, or are contaminated by extended
sources (2MASS XSC), or have proper motion ($pm > 3\,\sigma$ from
UCAC).  The resulting list contains about 3000 ISOCAM point sources
with $BRJHK$ and $LW2$ ($6.7\,\mu$m) photometry
(henceforth denoted ISOCP sources). The brightness ranges
are: $B = 14.5 \ldots 20.5\,$mag, $K = 12.5 \ldots 16\,$mag, $LW2 = 8 \ldots
12\,$mag $\pm 0.2 \ldots 0.4$ (Vega based system).
The typical errors on the $H - K_{\rm s}$ and $K_{\rm s} - LW2$  colours are 0.1
$\ldots$ 0.2 mag and 0.3 $\ldots$ 0.5 mag, respectively.

\vspace{-0.2cm}

\section{AGN candidate selection}
\begin{figure}
  \vspace{-0.3cm}
  \hspace{-0.65cm}
    \epsfig{file=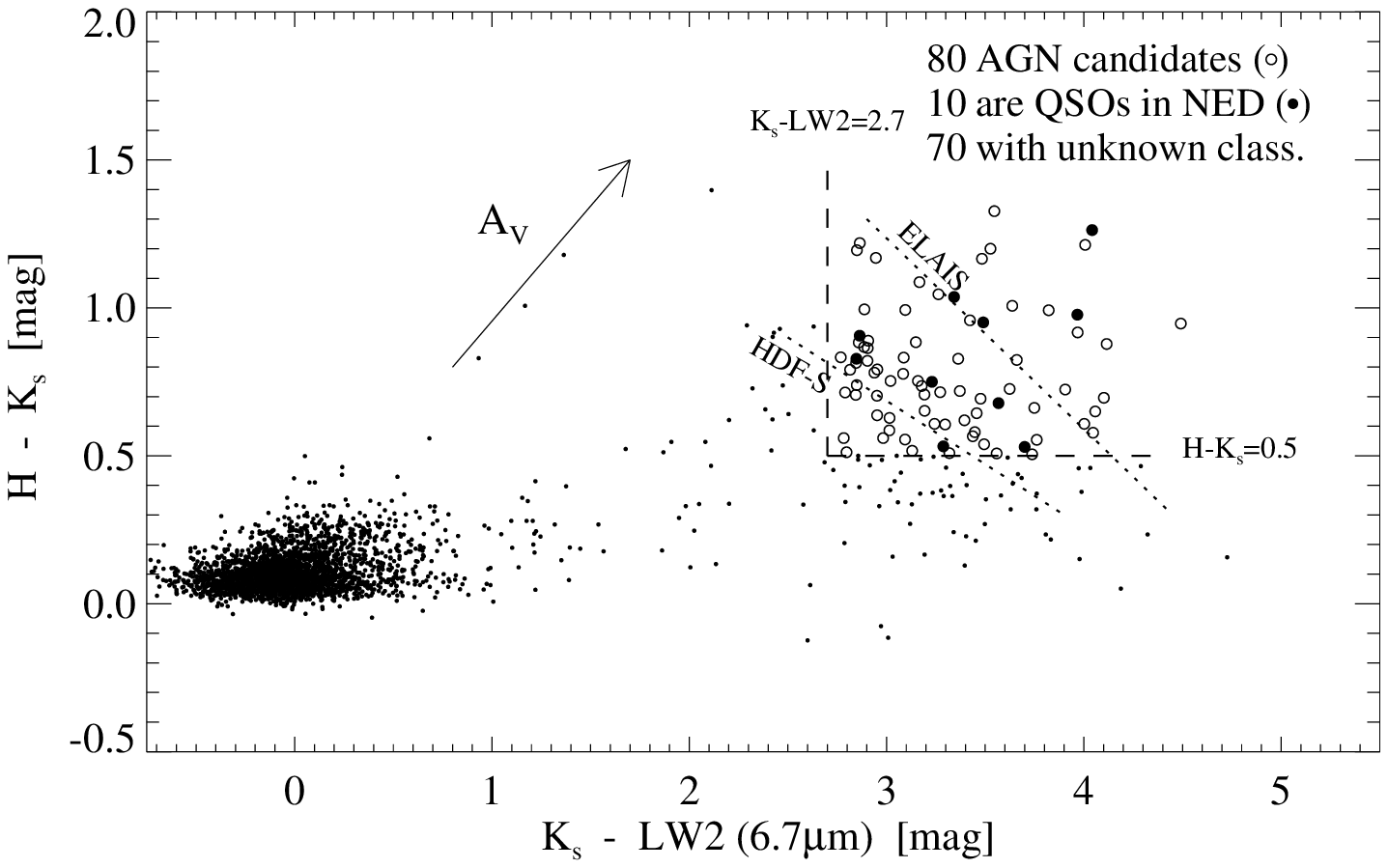,width=9.75cm,clip=true}
      \vspace{-0.5cm}
      
      \hspace{-0.65cm}
    \epsfig{file=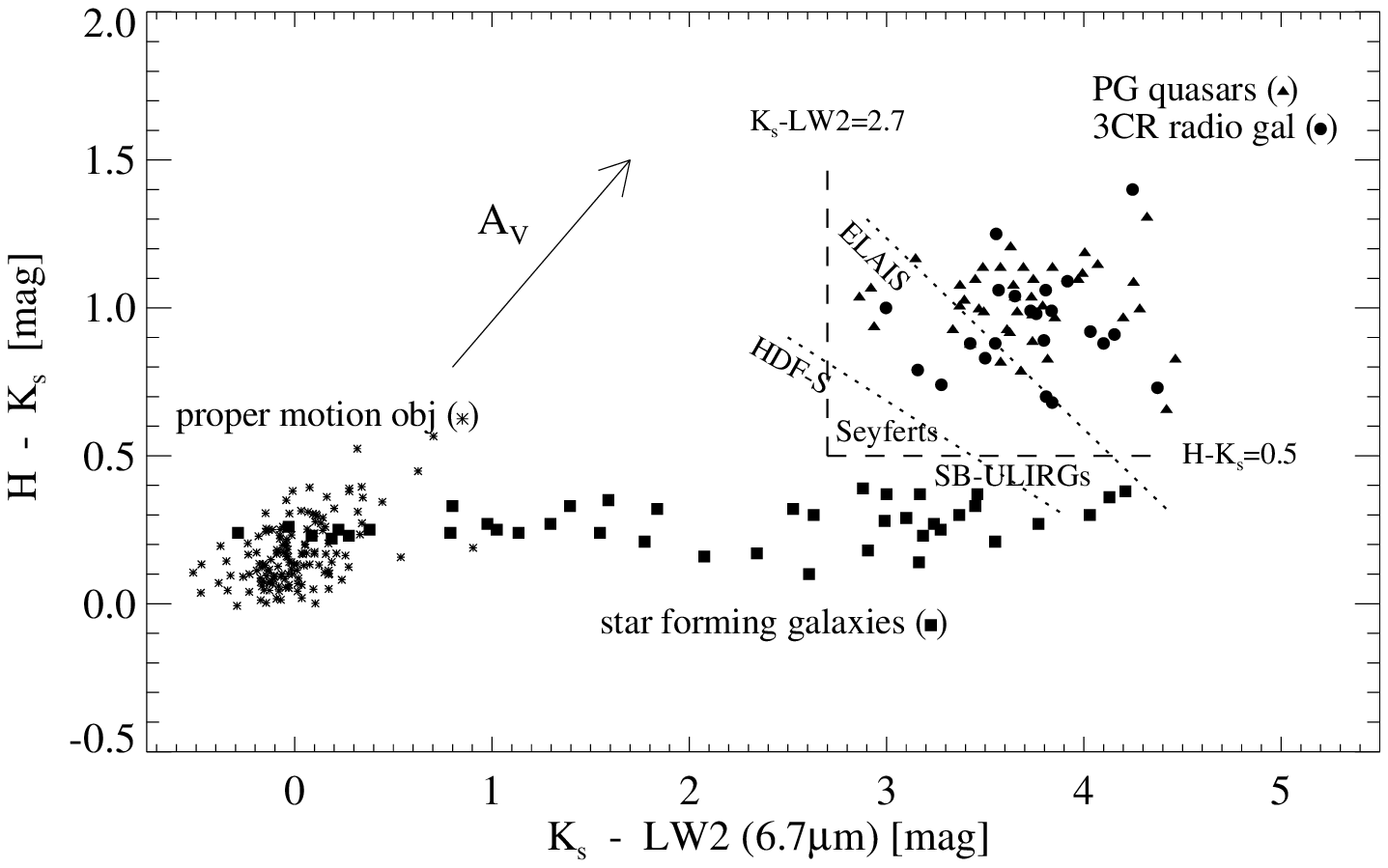,width=9.75cm,clip=true}
\caption[]{\label{msxxxx_fig1} {\it Top:} $H - K_{\rm s}$ vs.
$K_{\rm s} - LW2$ colour-colour diagram of the 3000 unresolved
ISOCAM sources at $|b| > 20^{\rm o}$. Circles and fat dots denote
the AGN candidates, small dots all other sources. The dotted
oblique lines indicate the limits reached by the HDF-S and ELAIS
sources, lying below and left hand of the lines. {\it Bottom:}
Comparison samples. }
 \vspace{-0.3cm}
 \vspace{-0.25cm}
\end{figure}

The ISOCP list contains various object classes like stars, normal
and active galaxies. Fig.\,\ref{msxxxx_fig1} (top) shows their
distribution in the $H - K_{\rm s}$ versus $K_{\rm s} - LW2$
colour-colour diagram. Noteworthy, the ISOCP sources clearly reach
more extreme IR colours than those of the Hubble Deep Field South
(Oliver et al. 2002, Mann et al. 2002) and 
the ELAIS fields  (V\"ais\"anen et al. 2002,
Rowan-Robinson et al. 2003). The corresponding limits are
indicated in Fig.\,\ref{msxxxx_fig1}.

To identify the types of objects that populate different regions
of the colour-colour diagram shown in Fig. 1 (top), we show the
same diagram for sources with known identification in Fig. 1
(bottom). Stars, i.e. proper motion objects, lie in the lower left
corner at $K_{\rm s} - LW2 < 1$ and $H - K_{\rm s} < 0.5$. This
area also contains radio-quiet elliptical galaxies from the
revised Shapley-Ames catalog.

UV-excess selected Palomar-Green (PG) quasars (AGN type 1, data
from Haas et al. 2003) and powerful FR\,2 radio galaxies from the
3CR catalogue (AGN type 2, data from Haas et al. 2004 and
Siebenmorgen et al. 2004) lie in the upper right corner at $K_{\rm
s} - LW2 > 2.7$ and $H - K_{\rm s} > 0.5$ -- a region that is also
populated by 3CR quasars (AGN type 1). As an additional check, we
cross-correlated the UV-selected quasars in the V\'eron-Cetty \&
V\'eron catalog (2003, 11th edition) with the 2MASS catalog. More
than 90\% of the quasars with  $V < 18$ mag and $z < 0.8$ lie at
$H - K_{\rm s} > 0.5$. At $z \sim 1.5$ the fraction goes down to
10\%, probably due to the shift of the NIR excess beyond the
$K_{\rm s}$-band. The sources from the PG and the 3CR
catalogs, respectively, shown in Fig.\,\ref{msxxxx_fig1} bottom
are at $z < 0.8$. For simplicity, we denote the sources in the
colour range $K_{\rm s} - LW2 > 2.7$ and $H - K_{\rm s} > 0.5$ as
excess sources.

The ISOCP excess sources are neither carbon stars (e.g.
Liebert et al. 2000) nor AGB stars (Joris Blommaert, priv. comm.),
both types showing optical and NIR colours different from these
ISOCP sources. Also, dusty young stars show different optical
colours and they are not expected at high galactic latitude $|b| >
20^\circ$.

Normal star forming galaxies (Boselli et al. 1998, Dale et
al. 2001) populate the full range of $K_{\rm s} - LW2$, reaching
into the regime of AGN, but they lie significantly lower in $H -
K_{\rm s}$ ($< 0.5$), so that they can be distinguished from the
PG/3CR AGN.  Even IR luminous starburst galaxies having  $10^{11}
< L_{\rm FIR}$ [L$_\odot] < 10^{12}$, like Arp\,157, NGC\,1569,
NGC\,2146, NGC\,3256 or M\,82, fall into this colour range ($H -
K_{\rm s} < 0.5$).

Seyfert 1 and 2 galaxies, from the sample by Clavel et al. (2001)
with averages listed in Tab.\,\ref{msxxxx_table_colours}, lie in
the "transition range" between normal star forming galaxies and
PG/3CR AGN. While Sy\,2s on average are below $H - K_{\rm s} =
0.5$, Sy\,1s are above and for clarity of the plot only the word
"Seyferts" delineates their colour location in
Fig.\,\ref{msxxxx_fig1} bottom.

\begin{table}
\caption[]{ Mean colours of Seyfert galaxies from Clavel et al. (2001)
  and ULIRGs from Klaas et al. (2001).  N gives the number of objects
  used for deriving the two colours, respectively.
\label{msxxxx_table_colours}
\vspace{-0.25cm}
}
\begin{center}
\footnotesize
\begin{tabular}{lcrcr}
              &  $H - K_{\rm s}$    &N   &    $K_{\rm s} - LW2$  &N \\
\hline				       		   		 	  
Sy1 ($<$1.8)  &  0.59  $\pm$   0.18 &25  &      2.8   $\pm$  0.6 &25\\
Sy2 ($\ge$1.8)&  0.39  $\pm$   0.18 &23  &      2.7   $\pm$  0.8 &23\\
				       		   		 	  
SBs-ULIRGs    &  0.49  $\pm$   0.17 &21  &      3.9   $\pm$  0.5 & 7\\
AGN-ULIRGs    &  0.94  $\pm$   0.35 & 5  &      3.9   $\pm$  0.4 & 5\\
\hline

\end{tabular}
\end{center}
\vspace{-0.5cm}
\vspace{-0.2cm}
\end{table}

%

The colours of IR ultra-luminous ($L_{\rm FIR} > 10^{12}\,$L$_\odot$)
starburst galaxies (including LINERs),
which we derived from the sample of Klaas et al.
(2001) together with 2MASS and ISOCAM 6.7$\mu$m data,
lie in the range $K_{\rm s} - LW2 > 2.7$ and  $0.2 < H - K_{\rm s} < 0.8$
(Tab.\,\ref{msxxxx_table_colours}).
13 of them lie below and 8 above $H - K_{\rm s} = 0.5$
overlapping with the range of the PG/3CR AGN.
In Fig.\,\ref{msxxxx_fig1} only the word "SB-ULIRGs" is placed at
their mean location.
On the other hand, the  AGN-ULIRGs
with "warm"  $F_{12}/F_{25}$ colours
(like Mrk\,231 and IRAS\,05189-2524)
clearly fall into the colour range of the PG/3CR AGN (Tab.\,\ref{msxxxx_table_colours}).

In following we discuss, whether the ISOCP excess sources could be
explained in terms of pure starburst IR galaxies. We will argue
that such starforming galaxies would be either too bright or too distant to
match {\it both} the observed fluxes at $K_{\rm s}$, $LW2$ and at
60$\mu$m {\it and} the corresponding colours of the ISOCP excess
sources. Our argumentation is essentially based upon the following
empirical facts:
\vspace{-0.2cm}
\begin{enumerate}

\item The ISOCP excess sources have median (mean) fluxes at
$K_{\rm s}$ and $LW2$ of 0.66\,mJy (0.9\,mJy) and 1.75\,mJy
(3.2\,mJy), respectively.

\item The $LW2$ flux of a starburst galaxy with $z<0.15$ is
mostly due to the PAH bands around 7.7\,$\mu$m (e.g. Laurent et
al. 2000). For
$z>0.15$ this feature is shifted out of the $LW2$ passband ($5 -
8.5\,\mu$m) resulting in a reduced  $K_{\rm s} - LW2$ colour below
2.7.

\item If a red ($H - K_{\rm s} > 0.5$) starburst galaxy has a
$K_{\rm s}-$band excess -- originating from hot dust close to the
sublimation temperature -- then its 60\,$\mu$m emission is also
expected to be strong. E.g. the five SB-ULIRGs observed by Klaas
et al. (2001) with $H-K_{\rm s} > 0.4$ show a flux ratio
$F_{60}/F_{LW2} \approx 230 \ldots 570$. Less luminous, more
gently star-forming galaxies might have a smaller $F_{60}/F_{LW2}$
ratio, but -- on the other hand -- the typical known SB-LIRGs like
M\,82 are bluer ($H-K_{\rm s} < 0.5$) than the present ISOCP
excess sources.

\end{enumerate}
\vspace{-0.2cm}

\textbf{PAH bands:}  First, we make use of the PAH bands and
discuss the ISOCP excess sources with $F_{2.2\,\mu{\rm m}} \la
1$\,mJy. In order to match this limit, a typical SB-ULIRG, like
e.g. IRAS\,17208-0014, with $F_{2.2\,\mu{\rm m}} \approx 20$\,mJy
at $z = 0.042$, would have to lie at $z = 0.188$. In this case,
however, $K_{\rm s}-LW2$ would attain values below 2.7, see point
(2.) above. The same argument applies to other SB-ULIRGs (like
MRK\,273, ARP\,220, IRAS\,14348-1447 and IRAS\,23365+3604). Hence,
a pure starburst counterpart of the {\it faint} ISOCP excess
sources must have a lower luminosity, i.e. being at most an
SB-LIRG.

{\boldmath $F_{60}/F_{LW2}$} \textbf{ratio:}  Second, we make use
of the $F_{60}/F_{LW2}$ ratio, which attains typically values
above 230 for SB-ULIRGs. A similar flux ratio is expected for
highly dust-enshrouded SB-LIRGs with $H-K_{\rm s}>0.5$ (down-sized
SB-ULIRGs), if they exist. In order to match $F_{LW2}>1\,$mJy for
the {\it bright} ISOCP sources, the expected flux at 60\,$\mu$m
would have to be at least $F_{60} \ga 230 \ldots 570$\,mJy, which
is above the IRAS detection threshold at low-cirrus high-galactic
latitudes.

For \emph{faint} ISOCP sources with fluxes $F_{LW2}$\,$<$\,1\,mJy
we expect useful IRAS upper limits. We have examined the 70
unclassified ISOCP excess sources
(Fig.\,\ref{msxxxx_fig1}, top) individually: only eight of them
show a marginal detection in the $60\,\mu$m IRAS ADDSCAN data, the
remaining sources have $3\sigma$ upper limits between 90 and
190\,mJy. Two of the detected sources exhibit
$F_{60}/F_{LW2}$\,$<$\,230,
all of the remaining sources have $F_{60}/F_{LW2}$\,$<$\,180 and
at least 40 of them even have $F_{60}/F_{LW2}$\,$<$\,100. These low
limits show that most of the 70 ISOCP sources are characterized by
a high MIR/NIR flux ratio, which is not accompanied by a high
FIR/MIR ratio as expected for known dust-enshrouded pure starburst
galaxies with $H-K_{\rm s}$\,$>$\,0.5.

If the ISOCP excess sources are not a new population of star-forming
galaxies, which are highly
dust-enshrouded (with red $H-K$), rich in PAH emission (with bright F$_{LW2}$)
and have cool FIR colours (with low 60$\mu$m flux), then the arguments
above favour a significant AGN contribution in these sources.
Furthermore, the four AGN-ULIRGs show an $F_{60}/F_{LW2}$ ratio
between 10 and 50, consistent with the 60$\mu$m upper limits found
for the ISOCP excess sources.

Based on the comparison with known object types we predict that
most of the 80 ISOCP excess sources house an AGN and have
Seyfert luminosities, but could be more powerful if at $z \ga
0.2$. Therefore, we consider the ISOCP excess sources as
promising AGN candidates. At redshift about $z > 0.8$, when the
AGN-typical SED bump shifts, the colours of the sources may become
bluer than $H - K_{\rm s} =0.5$, and a refined analysis using
other filters has to be applied in order to identify all AGN.
Remarkably, the new method is - \'a priori - not biased against
optical-UV selected QSOs.  For comparison, the 2MASS AGN search
catches only the range $H - K_{\rm s} > 0.7$ (roughly
corresponding to the $J-K_{\rm s} > 1.2$ criterion used by Francis
et al. 2004), hence may ignore more of the known AGN - even
among the local ones.

\begin{figure*}
 \vspace{-0.1cm}
 \hspace{-0.4cm}
  \epsfig{file=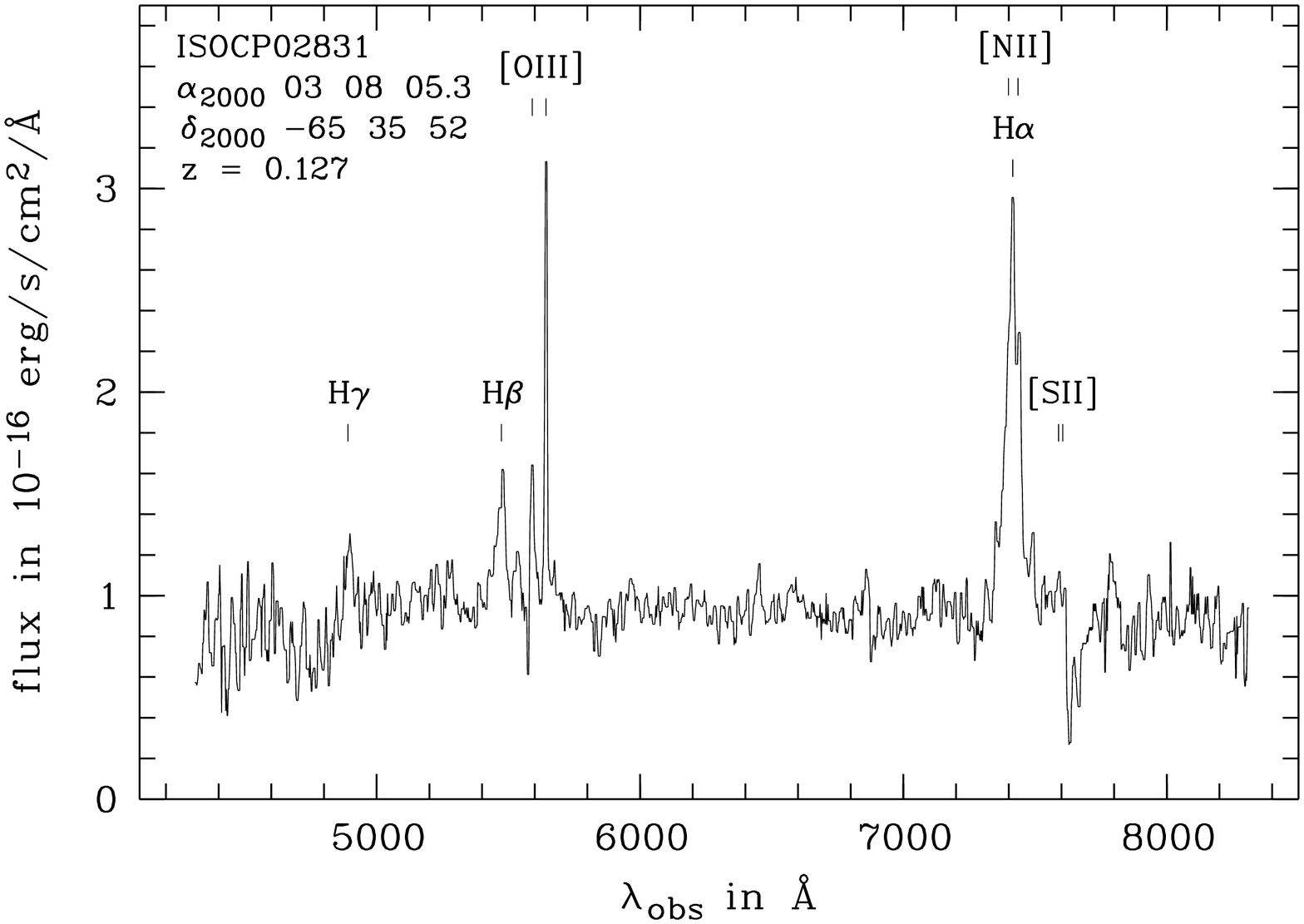,width=4.4cm,angle=270,clip=true}
 \hspace{-0.35cm}
  \epsfig{file=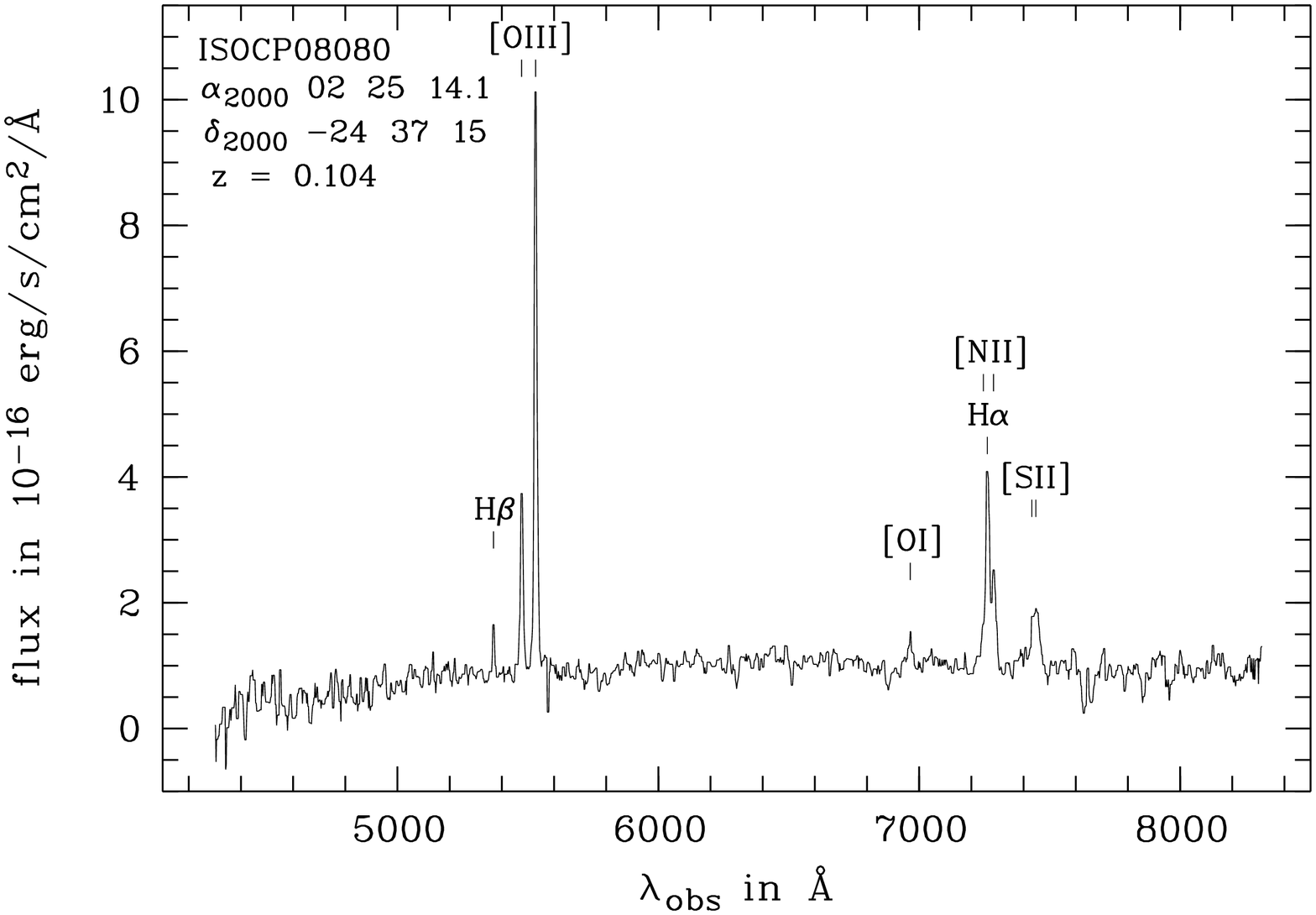,width=4.4cm,angle=270,clip=true}
 \hspace{-0.35cm}
  \epsfig{file=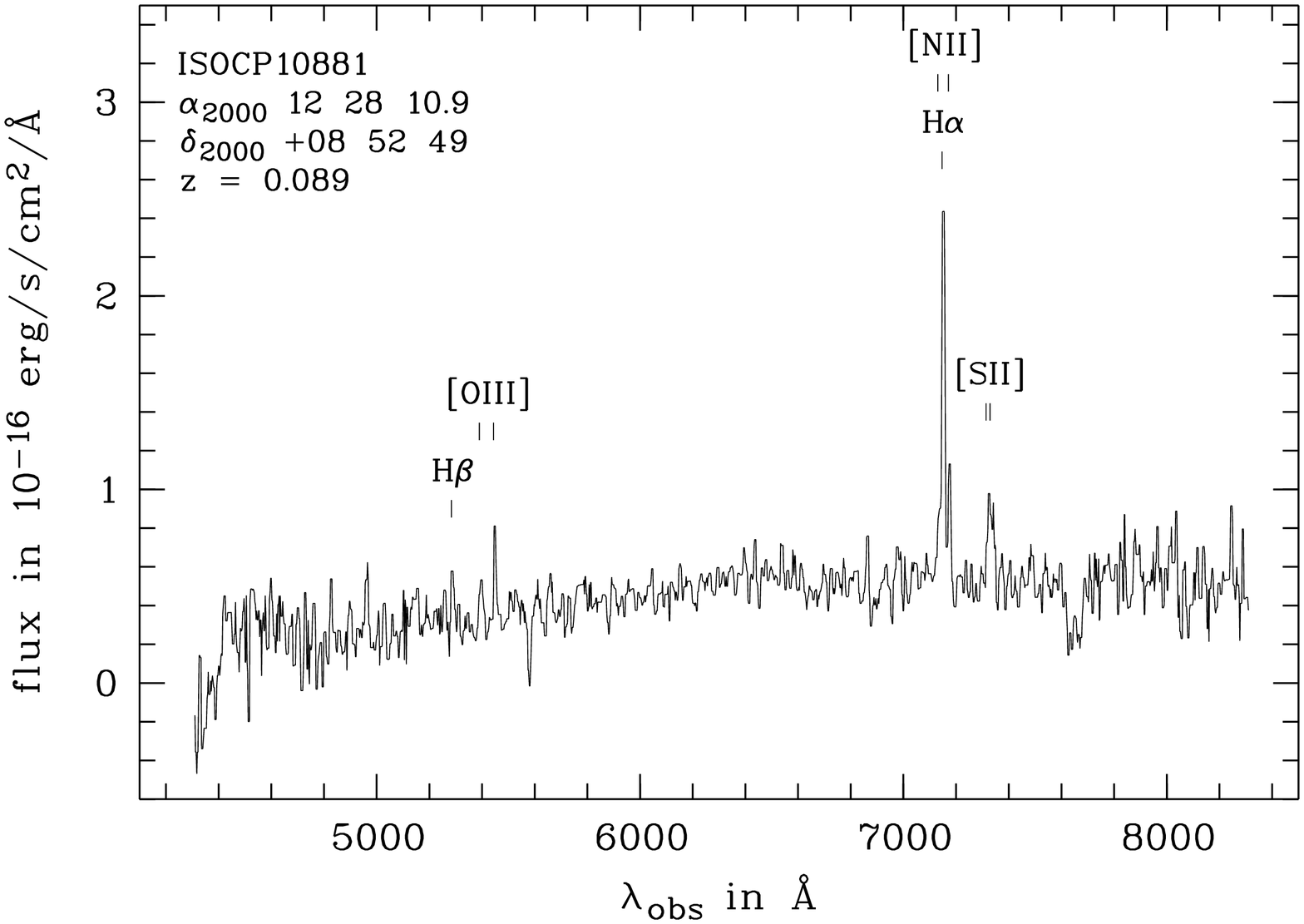,width=4.4cm,angle=270,clip=true}
 \vspace{-0.2cm}
\caption[]{
Optical spectra of three ISOCP sources with
$H - K_{\rm s} > 0.5$ and $K_{\rm s} - LW2 > 2.7$ obtained at the SAAO 1.9 m
telescope:
ISOCP02831: type-1 AGN with broad H$_{\alpha}$ and H$_{\beta}$ lines,
ISOCP08080: type-2 AGN,
ISOCP10881: reddened emission line galaxy with H$_\alpha$/H$_\beta >
10$.
\label{spectra}
}
\vspace{-0.4cm}
\end{figure*}

\vspace{-0.4cm}

\section{Optical spectroscopy}

Obviously, our predictions about the nature of the ISOCP excess
sources have to be verified by optical spectroscopy. Ten of our 80
AGN candidates are already listed in the NED as QSOs. None is
listed as a star (in SIMBAD) or starburst galaxy. Twelve sources have
1.4\,GHz radio detections, 4 being radio loud. Only one source
(3C345, Sandage \& Wyndham 1965) has also been identified
spectroscopically as QSO in the Sloan Survey (SDSS DR1, Abazajian
et al. 2003) while SDSS DR2 contains only one more spectrum of an ISOCP excess source
(RA,Dec$_{\rm J2000}$ = 13:25:07.8, +05:41:07) -- it is a QSO
at $z$\,=\,0.2.

Because of the marginal spectroscopic classification of the ISOCP
sources, we have started a spectroscopic survey for the 70 unknown
ISOCP excess sources ($K_{\rm s} - LW2 > 2.7$ and $H - K_{\rm s} >
0.5$).  First results obtained at the 1.9\,m SAAO telescope by end
of January 2004 on ten sources confirm their extragalactic nature
(Leipski et al. 2004 in prep.). The observations reveal two type-1
AGN and two type-2 AGN, one of which (ISOCP08080,
Fig.\,\ref{spectra}) has a high [\ion{O}{III}]$\lambda$5007
luminosity above 10$^{\rm  8}$ L$_{\odot}$ and can be classified
as QSO-2. Furthermore, we find two reddened LINER and four
extremely reddened emission-line galaxies. Example spectra are
given in Fig.~\ref{spectra}.  The visual extinction of the
emission-line galaxies as inferred from the ratio
H$_\alpha$/H$_\beta > 10$ is $A_V \ge 3\,$mag; actually $A_V$ is
expected to be much stronger when derived from data at longer
wavelengths, as found in many dust-enshrouded IR sources (e.g.
Haas et al. 2001). The redshift range is $ 0.1 \le z \le  0.5$.
Evidence is growing that LINER galaxies contain an AGN
(e.g. Satyapal et al. 2004).
Also the four emission-line galaxies may contain an AGN, since
their flux ratio $F_{60}/F_{LW2}$ $<$ 100 is below
that found for known pure starburst IR
galaxies. Therefore, the ongoing results with a fraction of 40\%
(4/10) optically identifyable classical AGN are consistent with
our prediction that most of the ISOCP excess sources are AGN.
Alternatively, we would have to postulate a new population of
dust-enshrouded unusually cool star-forming galaxies.

\vspace{-0.3cm}

\section{Conclusions}

We have discovered a sample of unique MIR excess sources. Various
arguments suggest that they likely contain an AGN. The ongoing
optical spectroscopy indicates 40\% of the sample to be classical
AGN and the remaining part to be dust-enshrouded sources. They
appear enigmatic in terms of having a high MIR/NIR flux ratio, but
relatively low FIR upper limits.  Since a moderate FIR/MIR flux
ratio is more typical for AGN, it  argues against pure starbursts.
This conjecture has to be confirmed. Because of the high
extinction, optical observations might not be able to see the true
AGN or starburst or composite nature of those sources. Therefore,
future observations with XMM-Newton, the VLA and IR spectroscopy
with the Spitzer Space Telescope are planned.

If all 80 ISOCP MIR excess sources turn out to be AGN and if 32 of
them (= 40\%) are classical AGN which could have been identified
also in optical surveys, then the number counts of AGN may
increase by a factor of up to 80/32 = 2.5 due to the inclusion of
dusty AGN.
Clearly, a reliable comparison will have to wait until spectra of
more ISOCP sources are obtained.

The emphasis of this letter is on presenting a promising strategy
how to find the presumed missing AGN. The selection of MIR excess
sources via the $H-K_{\rm s}$ and $K_{\rm s}-LW2$ colour-colour
diagrams in combination with the low 60$\mu$m upper flux limits
turns out to be efficient in separating starbursts from AGN.
It will establish a long sought-after technique for selecting the
full population of AGN largely free  of extinction.

\acknowledgements
We thank Michael Strauss and Nadia Zakamska
for their constructive and detailed referee report.
This research is  based on the Data Archives of ISO, 2MASS,
USNO-B, UCAC, IRAS, NVSS and FIRST.  The NED, SIMBAD and
the VisieR Service were used. This work is supported by the
Nordrhein-Westf\"alische Akademie der Wissenschaften, funded by the
Federal State and the Federal Republic of Germany.
BC thanks the South African National Research Foundation NRF for
financial support.

\end{document}